\documentclass[aps,prl,twocolumn,groupedaddress,showpacs]{revtex4-1}
\usepackage{graphicx}
\usepackage{booktabs}
\usepackage{amssymb}
\usepackage{amsmath}
\usepackage{multirow}
\usepackage{dcolumn}
\usepackage{bm}

\begin{document}



\title{First observation of the exchange of transverse and longitudinal emittances}


\author{J.~Ruan}
\email[]{ruanjh@fnal.gov} 
\affiliation {Fermi National Accelerator Laboratory, Batavia, IL 60510, USA}
\author {A.S.~Johnson} 
\affiliation {Fermi National Accelerator Laboratory, Batavia, IL 60510, USA} 
\author {A.H.~Lumpkin} 
\affiliation {Fermi National Accelerator Laboratory, Batavia, IL 60510, USA} 
\author {R.~Thurman-Keup} 
\affiliation {Fermi National Accelerator Laboratory, Batavia, IL 60510, USA}  
\author{H.~Edwards} 
\affiliation {Fermi National Accelerator Laboratory, Batavia, IL 60510, USA} 
\author {R.P.~Fliller} 
\altaffiliation {Current address: Photon Sciences Directorate, Brookhaven National Laboratory, Upton, NY 11973} 
\author {T.~Koeth} 
\altaffiliation {Current address: Institute for Research in Electronics and Applied Physics, University of Maryland, College Park, MD 20742} 
\author{Y.-E~Sun} 
\affiliation {Fermi National Accelerator Laboratory, Batavia, IL 60510, USA}


\date{\today}

\begin{abstract}
An experimental program to demonstrate a novel phase space manipulation in which the horizontal and longitudinal emittances of a particle beam are exchanged has been completed at the Fermilab A0 Photoinjector.  A new beamline, consisting of a $TM_{110}$ deflecting mode cavity flanked by two horizontally dispersive doglegs has been installed.  We report on the first direct observation of transverse and longitudinal emittance exchange: 
\{$\varepsilon_\text{x}^n$, $\varepsilon_\text{y}^n$, $\varepsilon_\text{z}^n$\}=\{$2.9\pm{0.1}$, $2.4\pm{0.1}$, $13.1\pm{1.3}$\}$\Rightarrow$\{$11.3\pm{1.1}$, $2.9\pm{0.5}$, $3.1\pm{0.3}$\} mm-mrad.
\end{abstract}


\maketitle

The next generation of advanced accelerators will benefit from the optimization of the phase-space volume by beam manipulations.  Such applications include high brightness light sources and improved luminosity for a linear $e^+$ / $e^-$ collider.  The advent of synchrotron radiation light sources and free electron lasers (FEL) has been a boon to a wide range of disciplines, resulting in a constantly increasing demand for brighter sources and better resolution \cite{bib:lcls}.  This demand translates to requirements on the properties of the underlying electron beams which produce the light.  In particular, one is driven to find ways to precisely manipulate the phase space volume of the beam to optimize it for the desired application \cite{bib:marie, bib:yine}.  It had been pointed out by Courant that while the total emittance (i.e. the phase space volume occupied by the beam) of a particle beam is conserved by a symplectic process, it does allow for the exchange of emittances between the 3 spatial dimensions \cite{bib:courant}.  Motivated by the FEL requirement for a small transverse emittance, Cornacchia and Emma developed a transverse / longitudinal emittance exchange (EEX) concept using a deflecting mode rf cavity located in the dispersive section of a magnetic chicane \cite{bib:emma}.  This method however, contained residual couplings between the two dimensions.  Other solutions exist that allow for complete exchange, such as the proposal by Kim to place a deflecting mode cavity between two magnetic doglegs \cite{bib:kim,bib:piot}.

In this Letter, we present the first experimental results of a near ideal, one-to-one exchange of transverse and longitudinal normalized emittances \cite{bib:emittance} at the Fermilab A0 Photoinjector (A0PI) using the latter scheme.  Unlike the original motivation which was to exchange a large incoming transverse emittance with a small incoming longitudinal one, this experiment exchanges a large longitudinal with a small transverse emittance.  There is however, no reason to expect that the opposite would not work as well.

The transfer matrix of the EEX beamline using thin lens elements for the dipoles and drifts and a thick lens cavity (symplectic) matrix for the 5-cell structure with the TESLA shape approximated by half-wavelength pillboxes is $M_\text{EEX}=$ 
\begin{equation}
\!\!
\left( \begin{array}{cccc}
0             & \frac{17\lambda}{40}                 & -\frac{1}{\alpha}-\frac{33\lambda}{40D}-\frac{L}{D} & -\frac{33\alpha\lambda}{40}-\alpha L \\
0             & 0                                                   & -\frac{1}{D}                  & -\alpha   \\
-\alpha       & -\frac{33\alpha\lambda}{40}-\alpha L               & \frac{17\alpha\lambda}{40D} & \frac{17\alpha^2\lambda}{40}        \\
-\frac{1}{D}  & -\frac{1}{\alpha}-\frac{33\lambda}{40D}-\frac{L}{D}     & \frac{17\lambda}{40D^2} & \frac{17\alpha\lambda}{40D}
\end{array} 
\right)
,
\end{equation}
where $\alpha$ is the bend angle of a dogleg, $L$ is the length of the drift, $\lambda$ is the wavelength and the cavity strength is set to $-1/D$, with D being the dispersion of a single dogleg \cite{bib:edwards}. 
In order to relate the final beam emittances to the initial, uncoupled emittances, we write the $4\times4$ beam covariance matrix $\Sigma_0$ whose elements are the average of the second central moments of phase-space variables $(x,x',z,\delta\equiv\frac{p_z}{\langle p_z \rangle}-1)$, 
\begin{equation}
\left( \begin{array}{cccc}
\langle x^2\rangle             & \langle xx'\rangle      & 0                        & 0 \\
\langle xx'\rangle             & \langle x'^2\rangle     & 0                         & 0   \\
0                              & 0                       & \langle z^2\rangle        & \langle z\delta\rangle \\
0                              & 0                       & \langle z\delta\rangle        & \langle \delta^2\rangle 
\end{array}\right)
,
\end{equation}
The beam matrix after traversing the EEX beamline is $\Sigma_{out}=M_{EEX}\Sigma_0M_{EEX}^T$.  The final rms emittances are found by taking the determinant of the $2\times2$ on diagonal sub-blocks of $\Sigma_{out}$ and can be written in terms of the incoming emittances as,
\begin{equation}
\!\!\!
\begin{array}{c}
\varepsilon_{x, \it out}^2 = \varepsilon_z^2 + (\frac{17\lambda^2}{40D})^2\langle x'^2\rangle\left[\langle z^2 \rangle + \alpha^2 D^2 \langle \delta^2 \rangle+2\alpha D \langle z\delta \rangle\right] \\ \\
\varepsilon_{z, \it out}^2 = \varepsilon_x^2 + (\frac{17\lambda^2}{40D})^2\langle x'^2\rangle\left[\langle z^2 \rangle + \alpha^2 D^2 \langle \delta^2 \rangle+2\alpha D \langle z\delta \rangle\right] \\
\end{array}
\!\!\!\!\!\!
\end{equation}
As can be seen, the non-zero cavity length causes an imperfect exchange which can, however, be reduced by proper selection of longitudinal or transverse input parameters \cite{bib:fliller, bib:ray}.

\begin{figure*}
 \begin{center}
 \includegraphics[width=6.4in]{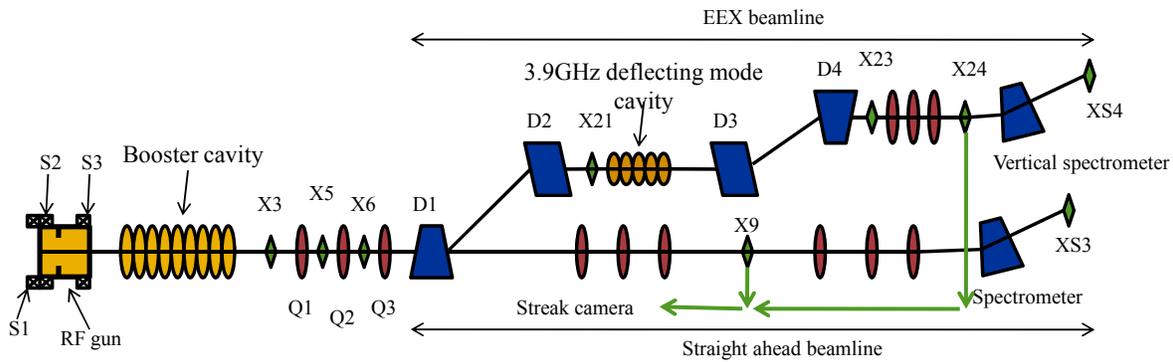}%
 \end{center}
\caption{\label{fig:beamline}Top view of the A0 Photoinjector showing elements pertinent to performing emittance exchange.
Elements labeled ``X'' are diagnostics stations (beam viewers and/or multi-slit mask locations), ``S'' are solenoid lenses,
``Q'' are quadrupole magnets and ``D'' are dipole magnets.}
 \end{figure*}

The A0PI facility includes an 1.5-cell normal-conducting L-band rf photocathode gun using a Cs$_2$Te photocathode irradiated by the frequency quadrupled, UV component of a Nd:Glass drive laser \cite{bib:carneiro}. The drive laser can be configured to provide a train of electron beam pulses separated by 1~$\mu$s with  charges up to 1~nC. Two emittance compensation solenoidal coils are installed as well as a bucking coil which is used to ensure zero magnetic field at the photocathode. The rf gun is followed by a 9-cell L-band superconducting cavity, and both a straight ahead and emittance exchange beam lines as schematically shown in Figure~\ref{fig:beamline}. 
 
The emittance exchange beamline at the A0PI consists of a 3.9~GHz $TM_{110}$ deflecting mode 5 cell cavity located between two horizontal dogleg magnetic channels.  The cavity is a liquid nitrogen cooled, normal conducting variant of a superconducting version previously developed at Fermilab \cite{bib:mcashan,bib:PAC07}.  The time varying longitudinal electric field gradient, $dE_z/dx$, of the $TM_{110}$ mode provides a linearly sloped field about the cavity axis.  The dispersion introduced by the first magnetic dogleg horizontally positions off-momentum electrons ($\delta \neq 0$) in the $TM_{110}$ cavity causing them to receive a negative longitudinal kick proportional to their $\delta$.  As a result, the $TM_{110}$  cavity reduces the momentum spread.   The time varying vertical magnetic field is $90\,^{\circ}$ advanced of the electric field.  The synchronous particle is timed to cross the cavity center at the peak of the electric field when the magnetic field is zero, and as a consequence, the cavity produces a time dependent positive (negative) horizontal kick with respect to early (late) particles.

Accurate measurements of the beam parameters are critical to the evaluation of the EEX process, thus the beamline is equipped with various diagnostic instruments.  Transverse beam profiles are measured by optical transition radiation (OTR) viewing screens oriented at $45\,^{\circ}$.  Both ingoing and outgoing transverse divergences are measured with the interceptive method of tungsten slits \cite{bib:wang}.  Downstream slit images are generated by single crystal YAG:Ce scintillator screens oriented orthogonal to the incident beam direction.  A $45\,^{\circ}$ mirror directs the radiation to the optical system.  This configuration eliminates depth of focus issues from the field of view and improves resolution \cite{bib:lumpkin-FEL}.  

\begin{figure}
 \includegraphics[width=3.2in]{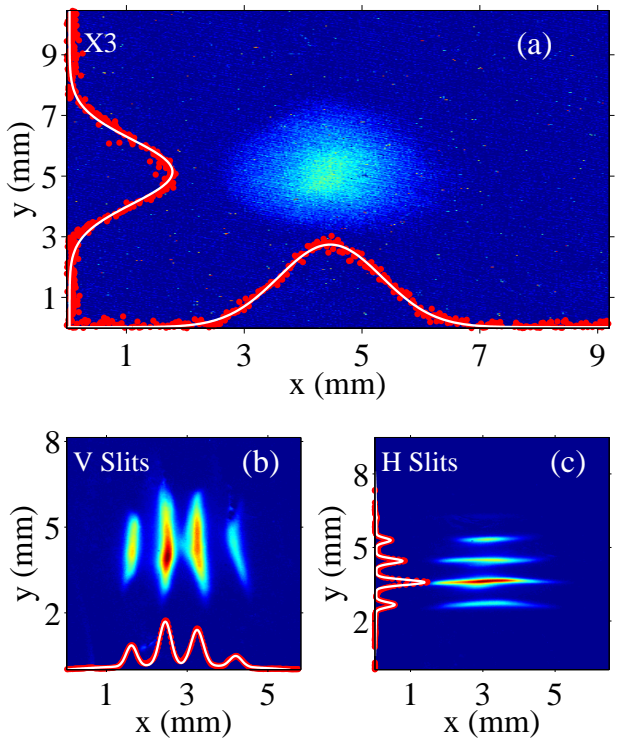}%
\caption{\label{fig:X03}Example incoming transverse emittance measurement data.  Figure (a) shows an OTR image of the beam spot at X3 with Gaussian fits to the projected x and y profiles.  Figures (b) and (c) are slit images taken at X6 YAG screen for x and y divergence measurements, respectively.  Gaussian fits to the projected profiles are shown.}
 \end{figure}

\begin{figure}
 \includegraphics[width=3.2in]{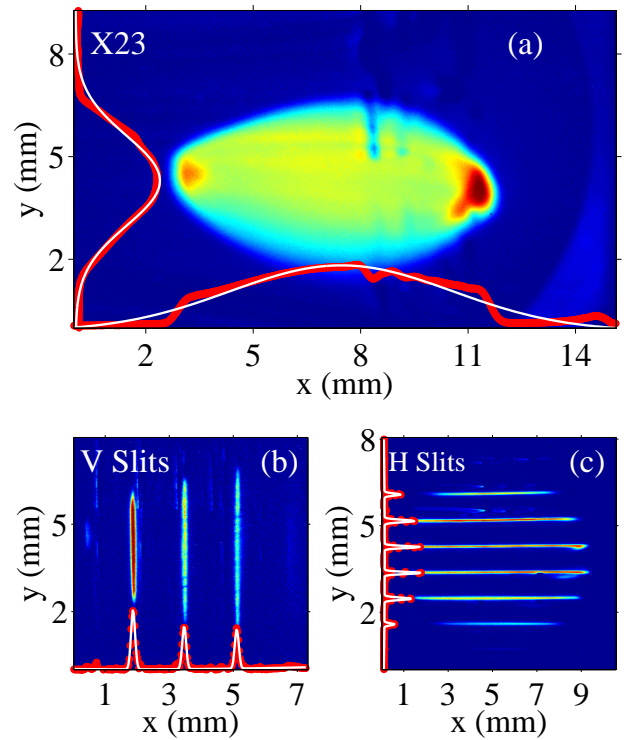}%
 \caption{\label{fig:X23}Example outgoing transverse emittance measurement data.  Figure (a) shows an YAG:Ce screen image of the beam spot at X23 with Gaussian fits to the projected x and y profiles.  Figures (b) and (c) are slit images taken at X24 YAG screen for x and y divergence measurements, respectively.  Gaussian fits to the projected profiles are shown.}
 \end{figure}

Example incoming beam and slit images are shown in Figure~\ref{fig:X03}.  The beam image is taken from the OTR screen located at X3.  Horizontal and vertical slits of 50~$\mu$m width separated by 1~mm are inserted into the beamline at X3, and the beamlets are allowed to drift 1.29~m to the YAG:Ce screen located at X6.  Image profiles are projected along the axis and fit with Gaussians.  Sample outgoing emittance measurements are shown in Figure~\ref{fig:X23}.  At X23 the horizontal slits are separated by 2~mm while the vertical slits are spaced at 1~mm.  A summary of input and output data is listed in Table~\ref{tab:parameters}.  Prior to image analysis, the dark current contributions have been subtracted by acquiring a background image with the beam shutter closed.  The uncertainty in the emittance includes the statistical fit uncertainty, pulse to pulse variation, and an estimate of the uncertainty in the optical resolution based on the differences between modulation contrast and edge blurring measurements using a calibration target.  A {\sc matlab}-based program calculates the emittances and the Courant-Snyder parameters ($\alpha$,$\beta$,$\gamma$) based on the X3-X6 and X23-X24 spot and slit image pairs.  Transverse beam position is monitored by $10$ button beam position monitors. 

\begin{figure}
\includegraphics[width=3.2in]{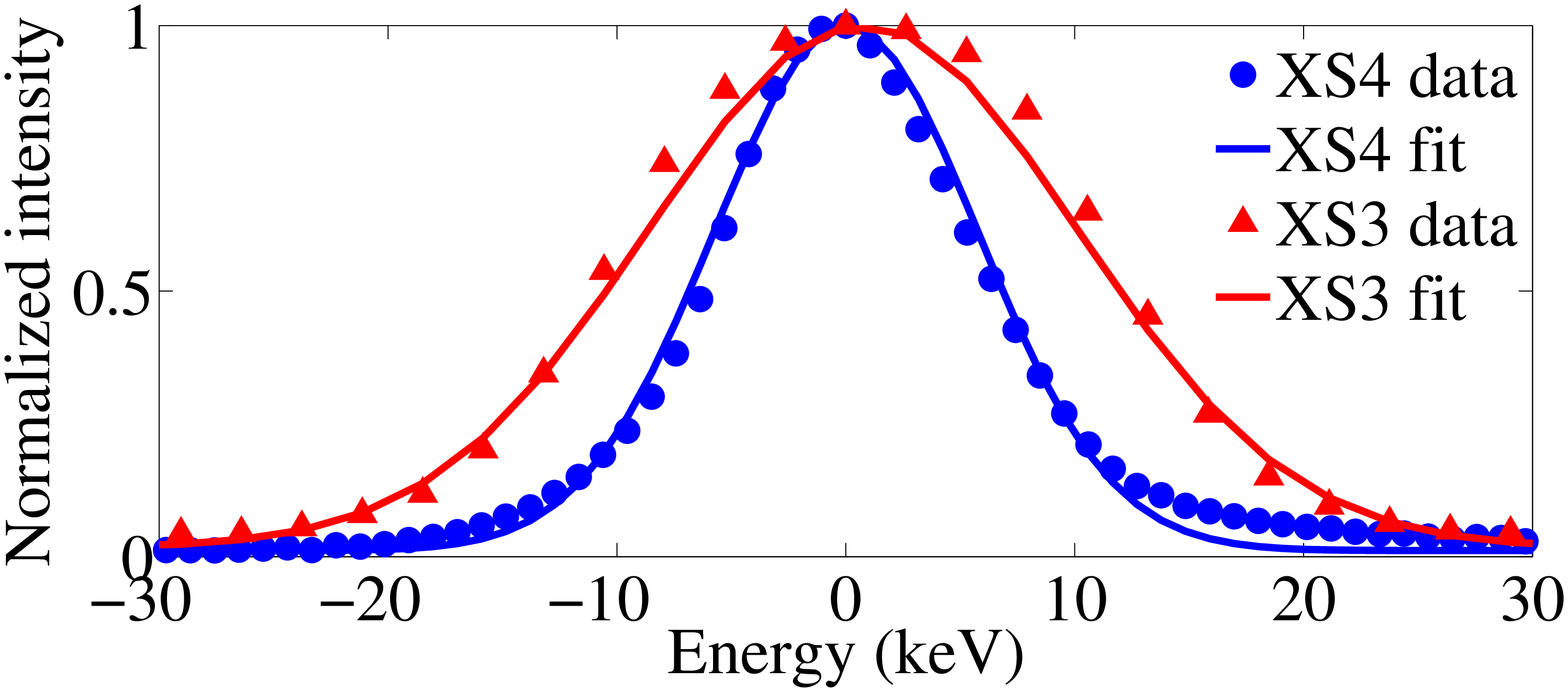}
\caption{\label{fig:ES}Energy spread measurements before and after EEX.  The triangles show typical incoming minimum energy spread as measured at XS3 with a Gaussian fit to the projection.  After EEX, the energy spread measured at XS4 is shown with dots and a Gaussian fit to the projection.}
\end{figure}

\begin{figure}
\includegraphics[width=3.2in]{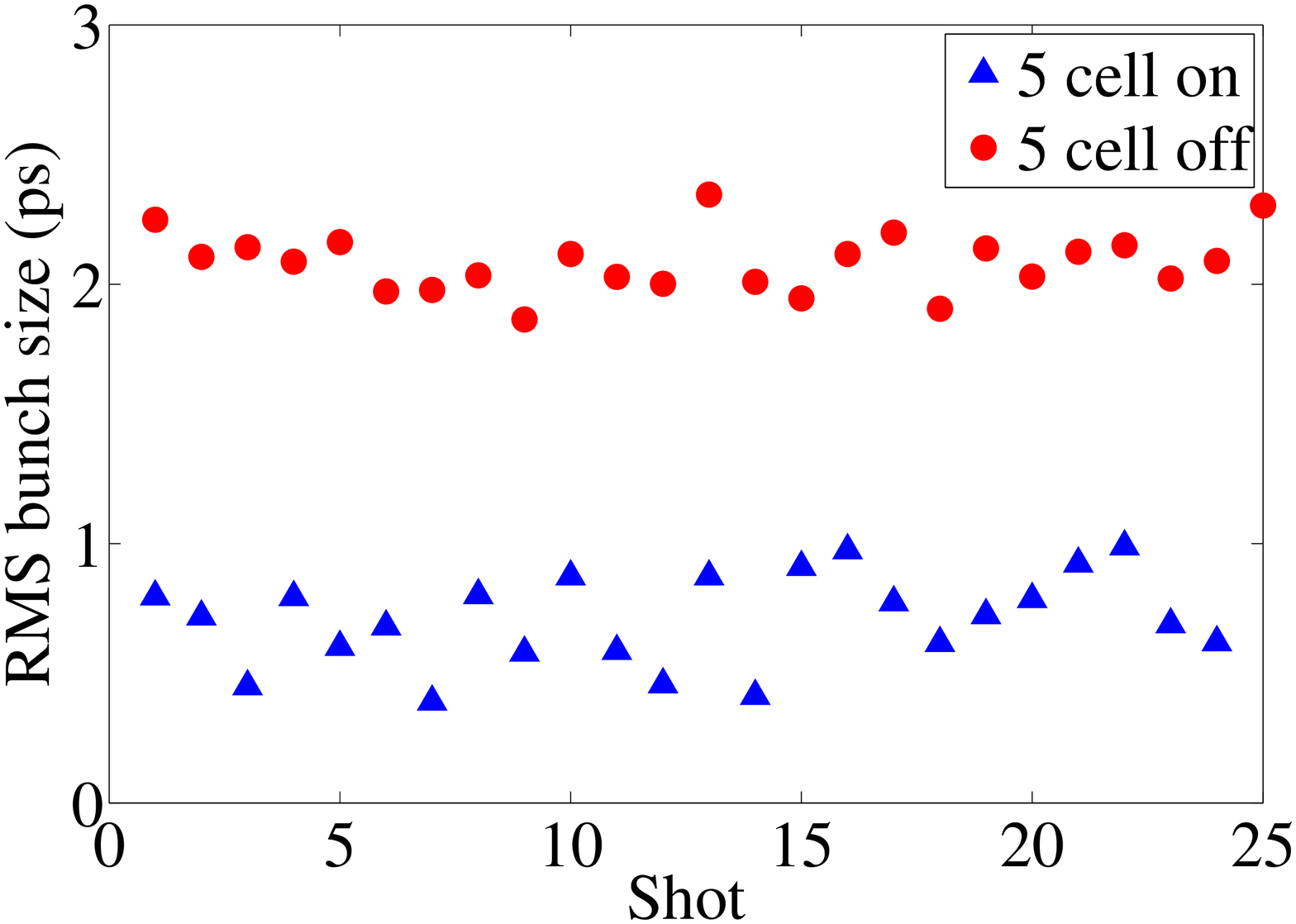}%
 \caption{\label{fig:length}Effect of deflecting mode cavity on bunch length.  The dots represent the bunch length as measured with the streak camera at X24 with the deflecting mode cavity off.  The triangles show a reduction in bunch length when measured with the deflecting mode cavity on.  Each measurement was made over 25 shots.}
 \end{figure}

Projected longitudinal emittance measurements are made by combining energy spread and bunch length measurements.  EEX input and output central momenta and momentum spreads are measured by two spectrometer magnets and down-stream viewing screens.  Figure~\ref{fig:ES} shows the energy spread with Gaussian fits as measured at XS3 and after EEX at XS4.  We conservatively report the output longitudinal emittance by only taking the energy-spread bunch-length product, $\varepsilon_\text{z,out}$ =  $\sigma_\delta\sigma_z$.  The bunch length is then determined at the X9 OTR screen using a Hamamatsu C5680 streak camera operating with a low jitter synchroscan vertical plug-in unit phase locked to 81.25~MHz as described previously \cite{bib:lumpkin}.  The outgoing energy-spread is measured at the XS4 screen following the vertical spectrometer magnet.  The bunch length measurement at X24 is made with OTR transported to the streak camera and with the far infrared coherent transition radiation transported to a Martin-Puplett interferometer \cite{bib:keup}.  As a graphic example of the effects on bunch length in the exchange process, Figure~\ref{fig:length} shows the effective compression by about a factor 3 with 5-cell cavity on (blue) compared to off (red).  

\begin{table}
\caption{\label{tab:parameters}
Summary of measured input and output rms beam parameters at 14.3 MeV with charge of 250 pC per bunch. 
}
\setlength{\tabcolsep}{9pt}
\begin{center}
\begin{tabular}{cr@{$\,\pm\,$}lr@{$\,\pm\,$}ll}
\hline
\hline
\toprule \toprule
Parameter & \multicolumn{2}{c}{In} & \multicolumn{2}{c}{Out} & Unit\\
\midrule
\hline
$\sigma_{x}$  & 0.905  & 0.013 & 4.014 & 0.059 & mm  \\
$\sigma_{x'}$  & 0.110  & 0.002 &  0.098 & 0.010  & mrad \\
$\sigma_{z}$  & 2.3  & 0.2 &  0.8 & 0.2 & ps  \\
$\sigma_{\delta}$ & 9.2  & 0.9 &  6.1 & 0.6  & keV \\
\bottomrule \bottomrule
\hline
\hline
\end{tabular}
\end{center}
\end{table}
\setlength{\tabcolsep}{6pt}  

\begin{figure}
\begin{center}
\includegraphics[width=3.2in]{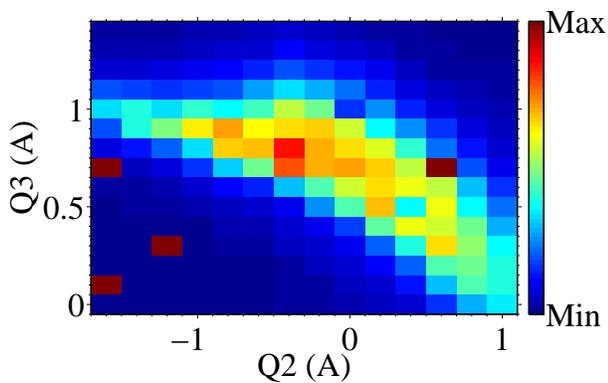} 
\end{center}
\caption{\label{fig:quad}
A relative output 1/$\sigma_\delta\sigma_z$ product map against input quadrupole currents.  
}
\end{figure}

The direct measurement of the emittance exchange has been performed at $\approx{14.3}$ MeV with a bunch charge of 250~pC, the latter chosen as a compromise between diagnostic requirements and space-charge effects.  To set up the incoming longitudinal phase space, the fractional momentum spread was minimized by operating the booster cavity off crest.  Separate experiments have shown the coherent synchrotron radiation (CSR) production at D3 is minimal at the selected 9-cell phase setting so we anticipate the emittance growth due to CSR is also low \cite{bib:charles}.  Input transverse parameters were tuned by adjusting Q1, Q2 and Q3 for a minimum EEX beamline output bunch-length energy-spread product, $\sigma_\delta\sigma_z$.  Since the intensity of the coherent transition radiation is strongly dependent on the bunch length, the interferometer's pyroelectric sensors are used to make quick, but uncalibrated, relative bunch-length measurements. This is very useful in mapping the effects of the input quadrupole fields on output longitudinal parameters.  A normalized 1/$\sigma_\delta\sigma_z$ product map is shown in Figure~\ref{fig:quad}.  Complete measurements of the initial and final emittances were collected with these conditions.  

For comparison, a linear transfer matrix model of the EEX beamline has been assembled in {\sc matlab} in an effort to explore the behavior of the EEX line.  It includes thick quadrupole and dipole magnets, and uses a thick lens model of the deflecting mode cavity composed of five zerolength $TM_{110}$ cavities each separated by a 3.9 GHz freespace halfwavelength drift, which agrees well with the realistic elliptical cavity transfer function \cite{bib:koeth}.  The measured emittance exchange transport matrix shows good agreement with the calculated transport matrix \cite{bib:PAC09}.

Results of the measurements are shown in Table~\ref{tab:results} and summarized as follows. The A0PI input beam's measured horizontal emittance is  $\varepsilon_\text{x}^n$=$2.9\pm{0.1}$~mm-mrad and the EEX output longitudinal emittance measured $\varepsilon_\text{z}^n$=$3.1\pm{0.3}$~mm-mrad demonstrating a 1:1 transfer of $\varepsilon_\text{x,in}^n$ to $\varepsilon_\text{z,out}^n$.  Similarly the input longitudinal emittance, $\varepsilon_\text{z,in}^n$=13.1$\pm{1.3}$~mm-mrad and the EEX output horizontal emittance measured $\varepsilon_\text{x,out}^n$=$11.3\pm{1.1}$~mm-mrad also show agreement between $\varepsilon_\text{z,in}^n$ and $\varepsilon_\text{x,out}^n$.  The vertical emittance was left unaffected, $\varepsilon_\text{y,in}^n$=$2.4\pm{0.1}$~mm-mrad $\Rightarrow$  $\varepsilon_\text{y,in}^n$=$2.9\pm{0.5}$~mm-mrad.  The combined results show the successful exchange of emittance between two planes while conserving the full 6D phase space volume.   
\begin{table}
\caption{\label{tab:results}
Comparison of direct measurements of horizontal transverse ($x$) to longitudinal ($z$) emittance exchange to simulation.  
Emittance measurements are in units of mm-mrad and are normalized.  
}
\setlength{\tabcolsep}{12pt}
\begin{center}
\begin{tabular}{rrrr@{$\,\pm\,$}lr@{$\,\pm\,$}lr@{$\,\pm\,$}l}
\hline
\hline
\toprule \toprule
  & \multicolumn{2}{c}{Simulated} & \multicolumn{4}{c}{Measured} \\
  & \multicolumn{1}{c}{In} & \multicolumn{1}{c}{Out} & \multicolumn{2}{c}{In} & \multicolumn{2}{c}{Out} \\
\midrule
\hline
$\varepsilon_\text{x}^{n}$  &  2.9  & 13.2  &  2.9  & 0.1 & 11.3 & 1.1   \\
$\varepsilon_\text{y}^{n}$  &  2.4  &  2.4  &  2.4  & 0.1 &  2.9 & 0.5   \\
$\varepsilon_\text{z}^{n}$  & 13.1  &  3.2  & 13.1  & 1.3 &  3.1 & 0.3   \\
\bottomrule \bottomrule
\hline
\hline
\end{tabular}
\end{center}
\end{table}
\setlength{\tabcolsep}{6pt}
In summary, a proof-of-principle transverse and longitudinal emittance exchange has been completed at the Fermilab A0 Photoinjector, demonstrating a novel fundamental phase space manipulation technique.  Further studies are planned at higher charge values to investigate the possible effects of space charge and CSR. 

\begin{acknowledgments}
We are grateful for the technical support of  J. Santucci, R. Montiel, W. Muranyi, B. Tennis, E. Lopez, C. Tan, M. Davidsaver, R. Andrews, B. Popper, G. Cancelo, B. Chase, J. Branlard and P. Prieto.  We greatly appreciate the discussions and comments from P. Piot (NIU), D. Edwards, M. Cooke, M. Stauffer and M. Cornacchia (UMD).  We thank M. Church, M. Wendt and E. Harms for their interest and encouragement.  This work was supported by Fermi Research Alliance, 
LLC under contract No. DE-AC02-06CH11359 with the U.S. Department of Energy.  
\end{acknowledgments}

\begin{thebibliography}{99}
\bibitem{bib:emittance}
The normalized emittance $\varepsilon^{n}=\gamma\beta\varepsilon$, where $\gamma$ is the Lorentz factor and $\beta=\sqrt{1-\gamma^{-2}}$ relates beams of different energies.
\end{thebibliography}

\end{document}